\documentclass[useAMS,usenatbib]{mn2e}

\usepackage{graphicx}
\usepackage[fleqn]{amsmath}

\newcommand{\mpl}{M_{\mathrm p}}
\newcommand{\im}{{\cal I}m}
\newcommand{\re}{{\cal R}e}
\newcommand{\rp}{r_{\mathrm p}}
\newcommand{\pp}{\varphi_{\mathrm p}}
\newcommand{\op}{\Omega_\mathrm{p}}
\newcommand{\okep}{\Omega_\mathrm{K}}
\newcommand{\cs}{c_\mathrm{s}}
\newcommand{\me}{\mathrm{M_\oplus}}
\newcommand{\xs}{x_\mathrm{s}}
\newcommand{\sgb}{\bar{\sigma}}
\newcommand{\hp}{h_\mathrm{p}}
\newcommand{\be}{ \begin {equation}}
\newcommand{\ee}{ \end {equation}}
\newcommand{\voc}[1]{{\bf{#1}}}

\title [Corotation torques and horseshoe drag]{On corotation torques,  horseshoe drag and 
the possibility of sustained stalled or outward protoplanetary migration}
\author[S.-J. Paardekooper and J.C.B. Papaloizou]{S.-J. Paardekooper$$\thanks{E-mail:
S.Paardekooper@damtp.cam.ac.uk} and J.C.B. Papaloizou\\
DAMTP, University of Cambridge, Wilberforce Road, Cambridge CB3 0WA, United Kingdom}
\begin{document}

\date{Draft version \today}

\pagerange{\pageref{firstpage}--\pageref{lastpage}} \pubyear{2008}

\maketitle

\label{firstpage}

\begin{abstract}
We study the torque on low mass protoplanets on fixed circular orbits, 
embedded in a protoplanetary disc  in the isothermal limit. 
We consider a wide range of surface density  distributions  including
cases where  the surface density increases smoothly outwards.
We perform both  linear disc response calculations and  non linear 
numerical simulations.  We consider a large
range of viscosities, including  the inviscid limit, as well
as a range of   protoplanet mass ratios, with special emphasis on the coorbital region and the corotation torque acting between disc and protoplanet.

For low mass protoplanets and large viscosity the corotation torque
behaves as expected from linear theory. 
However, when the viscosity becomes small enough to enable horseshoe 
turns to occur, the linear corotation torque exists only temporarily after
insertion of a planet into the disc,
 being  replaced by the horseshoe drag first discussed by Ward. 
This happens after a time that is  equal to the horseshoe libration period reduced by a factor  
amounting to about twice the disc aspect ratio.
 This torque scales with the  radial gradient of specific vorticity, as does the linear torque, 
but we find it to be   many times larger. 
If the viscosity is large enough for viscous diffusion across the coorbital region to occur 
within a libration period, we find that the horseshoe drag may be sustained. 
If not, the corotation torque saturates leaving only the linear Lindblad torques.
As the magnitude of the non linear coorbital torque (horseshoe drag) 
is always found to be larger than the linear torque, 
we find that the sign of the total torque may change even for for mildly positive surface density gradients. 
In combination with a kinematic viscosity large enough to keep the torque from saturating, strong sustained deviations from linear theory and outward or stalled migration may occur in such cases.
\end{abstract}



\section{Introduction}
A planet embedded in a gaseous disc is subject to a net torque that gives rise to orbital evolution. Since the discovery of the first hot Jupiter \citep{1995Natur.378..355M}, planet migration due to interaction with the protoplanetary disc has become a necessary ingredient of planet formation theory. A considerable amount of analytical and numerical work has gone into understanding the disc-planet interactions leading to planet migration \citep[see][for an overview]{2007prpl.conf..655P}.   

Three modes of migration can be distinguished, in most cases leading to migration towards the central star. High mass planets, for which the Hill sphere is larger than the disc scale height, open up deep gaps in the disc, after which migration proceeds on a viscous time scale \citep{1986ApJ...309..846L}.  For standard disc parameters, planets more massive than Jupiter migrate in this Type II regime \citep{2007MNRAS.377.1324C}. Intermediate mass planets, comparable to Saturn, embedded in massive discs may undergo runaway or Type III migration \citep{2003ApJ...588..494M}, which may be directed inward as well as outward \citep{2008MNRAS.387.1063P}. In this paper, we focus on low mass planets, with masses typical up to a few times the mass of the Earth ($\me$). Low mass planets excite linear waves in the disc, the action of which leads to Type I migration \citep{1997Icar..126..261W}. Linear, semi-analytical calculations have resulted in a widely used torque formula \citep[][hereafter TTW02]{2002ApJ...565.1257T} for isothermal discs.  

The total torque on an embedded planet can be decomposed in torques due to waves, which are generated at Lindblad resonances and propagate away from the planet, and corotation torques, generated near the orbit of the planet, where material on average corotates with the planet \citep[see][]{1979ApJ...233..857G}. For low mass planets, the wave torque is usually thought to dominate the total torque, and, being relatively insensitive to background gradients, lead to inward migration.
   
Although Type I migration was thought to be mathematically well-understood, it nevertheless posed a serious problem for planet formation. Applying the torque formula from TTW02 \citep[see also][hereafter KP93]{1993Icar..102..150K} to planets of a few $\me$ embedded in a typical disc resulted in inward migration time scales that are much shorter than the lifetime of the disc. In other words, these planets would all migrate to very small orbital radii, or even into the central star \citep{1997Icar..126..261W}. Since gaseous giant planets are thought to form around a solid core that is in this mass range, Type I migration theory essentially predicts that there should be no planets at large radii. This has led to investigations on how to slow down or stop Type I migration, for example through the action of magnetic fields \citep{2003MNRAS.341.1157T,2004MNRAS.350...849N} or sharp surface density gradients \citep{2006ApJ...642..478M}.

More recently, efforts have been made to relax the isothermal assumption that was made in previous works, and to properly account for the energy budget of the disc. Since radiation is the dominant cooling agent, radiation-hydrodynamical simulations are needed. The outcome of these simulations were surprising: \cite{2006A&A...459L..17P} found that the torque on the planet was \emph{positive}, leading to outward migration, whenever the opacity of the disc at the location of the planet was high enough to make cooling inefficient. In \cite{2008A&A...478..245P} it was recognised that this positive torque was a corotation effect, related to a radial entropy gradient in the unperturbed disc.  \cite{2008ApJ...672.1054B}, through a linear analysis of the corotation torque, suggested that the linear corotation torque in the presence of an entropy gradient can be strong enough to overcome the negative wave torque. However, \cite{2008A&A...485..877P} showed that this linear contribution is small, and that a genuinely non-linear effect is responsible for the change of sign of the total torque. 

In this paper, we take a step back and reanalyse the isothermal case, in a two-dimensional set-up. The simplicity of this model allows us to perform numerical
simulations with high enough resolution investigate in detail non-linear effects on the corotation torque
and to run them for long enough to study its possible saturation. 
We will show that linear theory only applies  for short times
on the order of a few orbits when  a protoplanet is inserted into a disc
for which the viscosity is not too large and there is a non-zero corotation torque.
 We also show that it is a non-linear effect associated with horseshoe bends \citep{1991LPI....22.1463W} that results in a departure from linear theory. This departure is found to
 result in torques that can have a much larger magnitude than the linear ones. 
 As the torque scales with the gradient of specific vorticity, this may produce noticeable effects
 that may even lead to stalled or outward migration, when this gradient is relatively large
 as for example occurs when the surface density increases gently outwards. For these effects to be sustained
 the viscosity must be large enough to prevent torque saturation. Viscosities comparable to those often 
 assumed in protoplanetary disc modelling are found to be large enough to enable these non linear
 corotation torques to be sustained for protoplanets in the Earth mass range.

The plan of the paper is as follows. We start in Section \ref{secEq} by reviewing the model used throughout the paper. In Section \ref{secCoro}, we review and discuss  linear corotation torque estimates
 as well as torque estimates based on  the horseshoe drag experienced by material
 executing horseshoe turns. These are shown to be distinct phenomena
 with different dependencies on the physical variables, the horseshoe drag
 being essentially non linear even though the associated torque scales in the same way as the linear one.
  We go on to  perform detailed linear calculations in Section \ref{secLin}, and then compare the linear and non-linear (horseshoe drag) torques in Section \ref{secNonlin}. In Section \ref{secNum}, we present the results of numerical hydrodynamical simulations, torques are also
  compared to linear and horseshoe drag estimates confirming
  the view outlined above. The horseshoe drag is found to be  significantly larger than the
linear corotation torque and potentially  a very important contributor
to the total torque for disc surface densities that increase even mildly outwards.
 We go on to consider the long term
  torque evolution and saturation as a function of viscosity finding corotation
  torque saturation in low viscosity cases and sustained corotation torques
  when the viscosity is large enough to resupply angular momentum to the coorbital region.
  We  discuss our results further in Section \ref{secDisc}.
   Finally we present a short summary, together with some concluding remarks, in Section \ref{secCon}. 
   In addition  we give an analysis and discussion of the time development of the linear  corotation 
   torque acting on a planet after immersion into a disc, demonstrating that the characteristic
   time is the orbital period as was found in the numerical simulations.

\section{Basic equations}
\label{secEq}
The evolution of a gaseous disc is governed by the Navier-Stokes equations. We will work in a cylindrical coordinate polar frame $(r,\varphi)$, centred on the central star, and we integrate the equations of motion vertically to obtain a two-dimensional problem. We are then left with the continuity equation and two equations of motion:
\begin{eqnarray}
\frac{\partial \Sigma}{\partial t}+\nabla \cdot \Sigma \voc v=0\label{eqcont}\\
\frac{\partial \voc v}{\partial t}+(\voc v \cdot \nabla)\voc v=-\frac{\nabla p}{\Sigma}-\nabla \Phi + \voc{f}_\mathrm{visc}\label{eqmot},
\end{eqnarray} 
where $\Sigma$ is the surface density, $\voc v=(v,r\Omega)^T$ is the velocity, $p$ is the vertically integrated pressure, $\Phi$ is the gravitational potential and $\voc{f}_\mathrm{visc}$ represents the viscous force. We use a locally isothermal equation of state, $p=\cs^2 \Sigma$, where the sound speed $\cs$ may be a prescribed function of radius. We take $\cs^2$ to be a power law with index $-\beta$, which makes $-\beta$ basically the power law index of the radial temperature profile. Writing $\cs=H\okep$, where $H$ is the vertical pressure scale height and $\okep$ is the Keplerian angular velocity, we usually adopt either a sound speed that gives rise to a constant aspect ratio $h=H/r$ (or, equivalently, $\beta=1$), or a purely isothermal disc with constant $\cs$ ($\beta=0$). In the latter case, we usually quote the aspect ratio at the location of the planet, $\hp=\cs/(\rp\op)$. The initial, or for linear
calculations the background,  surface density is taken to be a power law with index $-\alpha$.  Thus
$\Sigma = \Sigma_\mathrm{p}(r/\rp)^{-\alpha},$ where $\Sigma_\mathrm{p}$
is the surface density at the location of the planet.
The gravitational potential $\Phi$ is the sum of the potential due to the central star, the planet's potential $\Phi_\mathrm{p}$, and an indirect term that arises due the the acceleration of the coordinate frame centred on the star. For $\Phi_\mathrm{p}$, we use a softened point mass potential:
\begin{equation}
\Phi_\mathrm{p}=-\frac{G\mpl}{\sqrt{r^2-2r\rp\cos(\varphi-\pp)+\rp^2 +b^2\rp^2}},
\end{equation}
where $G$ is the gravitational constant, $\mpl$ is the mass of the planet, and $(\rp,\pp)$ are the coordinates of the planet. We will also use $q$ to indicate the mass ratio $\mpl/M_*$, where $M_*$ is the mass of the central star. The softening parameter $b$ should be a sizeable fraction of $h$, in order to approximate the result of appropriate vertical averaging of the potential. 

The exact form of the viscous terms can be found
 elsewhere \citep[e.g.][]{2002A&A...385..647D}. 
We take the kinematic viscosity $\nu$ to be a power law in radius,
 such that the initial surface density profile is a stationary solution.
 It is easy to see that the required viscosity law is 
$\nu \propto r^{\alpha-1/2}$. 
This way, we can neglect a global radial velocity field in the model, 
which greatly simplifies the analysis. 
Values quoted in the text refer to $\nu$ at the location of the planet,
 and will be in units of $\rp^2\op$, where $\op$
 is the angular velocity of the planet.
 

\section{Corotation torques}
\label{secCoro}
The wave (or Lindblad) torque exerted by a low-mass planet on a 
gaseous disc is relatively well-understood
 \citep[][TTW02]{1979ApJ...233..857G}. When the Hill sphere
 of the planet is much smaller than the scale height of the disc,
 the density  waves are excited at Lindblad resonances
 are  linear and the resulting torque
 can be calculated by  performing a linear 
response calculation and then summing the contributions  arising
from individual  Fourier components (TTW02).
 This torque usually leads to inward migration,
 and has  been thought to dominate in the Type I regime of low mass planets.
There are two approaches that can be followed in order 
to obtain  corotation torques.
The first, which we will refer to as the linear estimate, is based
on a linear response calculation. The second is a fundamentally
different approach based on  a direct torque
calculation made after making some assumptions about the 
trajectories of the disc  fluid elements, 
which we refer to as  a horseshoe drag calculation. 
The first approach applies at early times after a protoplanet is inserted into a disc
as, at least for a low mass protoplanet, 
 the response must first of all be linear. The second approach applies
at later times after the streamline pattern has adjusted to the
presence of the planet.
 
\subsection{Formula for the linear corotation torque for each azimuthal
mode number}
  In order to obtain linear estimates for corotation torques,
 the equations are linearised (see Section \ref{secLin}) and Fourier-decomposed in azimuth. 
The resulting single second-order equation \citep[see][]{1979ApJ...233..857G} 
  governing the disc response  can be solved to find the 
 corotation torque  acting on the planet.
Using an approximation scheme that assumes the perturbing
potential varies on a scale significantly longer then the scale height,
\citet{1979ApJ...233..857G} find this torque given by
\begin{equation}
\label{eqTmlin}
\Gamma_{\mathrm{c},m}=-\frac{m\pi^2 \Phi'^2_m}{2d\Omega/dr}\frac{d}{dr}\left(\frac{\Sigma}{\omega}\right),
\end{equation}
where $\Phi'_m$ is the $m$-th Fourier component of the planet's potential, 
$m$ is the azimuthal mode number and $\omega$ is the flow vorticity, being equal to $\Omega/2$ in a Keplerian disc.  
All quantities in equation (\ref{eqTmlin}) should be evaluated at corotation $(r = \rp)$,
which makes the total torque proportional to the radial gradient 
in specific vorticity, or vortensity, in the unperturbed flow there. 
Of course unless the softening parameter is rather large,
the perturbing potential does not vary slowly on the corotation circle
and so (\ref{eqTmlin}) cannot be used as described above.
However, it   becomes valid if the perturbing
potential ${ \Phi'_m}$ is replaced by the generalised potential
obtained by adding the enthalpy perturbation to it
\citep[see e.g.][and the appendix]{2006MNRAS.368...917Z}. 
But then the linear response equations
need to be solved in order to determine the enthalpy perturbation in order
to evaluate torques using  the modified form of (\ref{eqTmlin}),
which are nonetheless still proportional to the gradient
of specific vorticity.

\subsection{The total linear corotation torque in the limit of zero softening}
To obtain the total torque, one needs to sum the contributions
from  all values of $m$ 
\citep[see][]{1989ApJ...336..526W}.
For a Keplerian disc with zero softening,
 the total torque has been calculated, by finding the linear response numerically,
  to be given by
(see TTW02)
\begin{equation}
\label{eqTlin}
\Gamma_\mathrm{c,lin}=1.36\left(\frac{3}{2}-
\alpha\right)\frac{q^2}{\hp^2}\Sigma_\mathrm{p}\rp^4\op. \end{equation} 
For a 3D  disc it is found that the same expression holds
but with the numerical coefficient being replaced by $0.632.$

An important issue is the time required to develop the
linear corotation torque. Being linear this should not
depend on the mass of the protoplanet but only on intrinsic disc parameters.
Furthermore when a protoplanet is inserted into a disc, there has to be
an initial linear phase and for sufficiently low protoplanet mass,
the full linear response should develop. As it is somewhat involved, we
relegate the discussion of the time development of the linear corotation torque to 
the Appendix. From this discussion, the characteristic development time
is  expected to be on the order of the orbital period.
We now go on to consider the subsequent development of
the corotational flow.

\begin{figure}
\centering
\resizebox{\hsize}{!}{\includegraphics[]{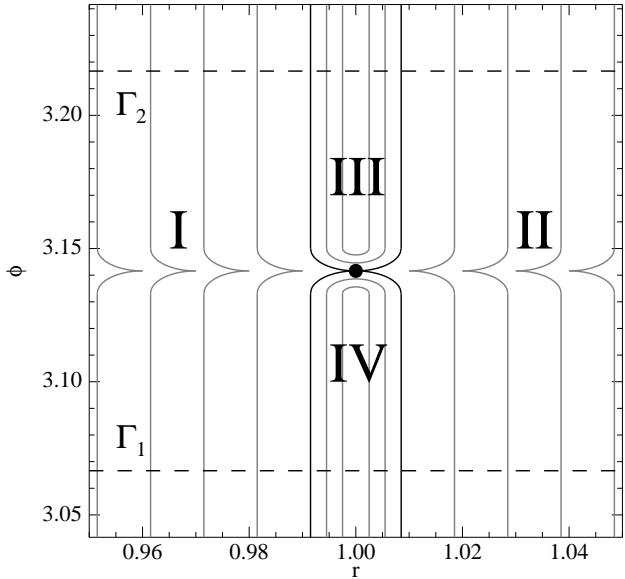}}
\caption{Schematic overview of the horseshoe region, with the planet denoted by the black circle. Solid grey lines indicate approximate streamlines, with the separatrices in black. The separatices divide the region into four parts, labelled by Roman numerals: I and II, the inner and the outer disc, respectively, and III and IV the leading and trailing part of the horseshoe region, respectively. Two horizontal dashed lines indicate the boundaries of the torque calculation, $\Gamma_1$ and $\Gamma_2$. Note that in this figure, the location of $\Gamma_1$ and $\Gamma_2$ is chosen arbitrarily. In the model, they are chosen to be far enough from the planet so that the corotation torque is determined within.}
\label{fighorseschem}
\end{figure}

\subsection {The horseshoe drag}
\cite{1991LPI....22.1463W} derived a formula for the corotation torque 
using a fundamentally different approach. The 
 argument was  based on the expected form 
of the gas streamlines. These can be classified into four groups as viewed
in a reference frame corotating with the planet (see Fig. \ref{fighorseschem}).
The first pass the protoplanet  interior to the coorbital region and  the second
pass  it exterior to the coorbital region. 
These groups extend to large distances from the protoplanet.
The other two groups consist of
material in the corotation region on horseshoe orbits that  execute turns close to the planet.
The third group approach and leave the protoplanet from 
 its  leading side while  the fourth do so from its  trailing side.
These groups are separated by two separatrices
 \citep [see e.g..][for more discussion of this aspect]{2006ApJ...652..730M,2008A&A...485..877P}.
For low mass protoplanets, most of the corotation torque is produced in a region
close to the planet with a length scale expected to be the larger of $b\rp$ or $H.$
It is important to note that it takes some time for the streamline structure
described above to develop  on this scale  after insertion of a protoplanet into a disc.
 As  this structure represents a finite deviation from the initial form, this  time
depends on the mass of the protoplanet. 
The calculation of the horseshoe drag applies to the situation when this streamline structure
has developed on this scale. It is important to note that the time required
is shorter  than that required to develop it on a scale comparable to $\rp$
which is when issues of torque saturation need to be considered.

Following \cite{1991LPI....22.1463W} we consider the torque produced by material
on streamlines undergoing horseshoe turns. We consider a region ${\cal R}$  interior to the two
separatrices separating these from the first and fourth group of streamlines that is bounded by two
lines of constant $\varphi,$ 
 $\Gamma_1$ and $\Gamma_2$ on the trailing and leading sides  of the protoplanet
respectively (see Fig. \ref{fighorseschem}). 
These boundaries are supposed to be sufficiently far from the protoplanet
that the corotation torque is determined within. 
Assuming a steady state this torque may be obtained by
considering the conservation of angular momentum within ${\cal R}$ written in the form

\begin{eqnarray}
\Gamma_\mathrm{c,hs}  =  \int_{\cal R}  
\Sigma\left( {\partial \Phi_\mathrm{Gp}\over \partial \varphi }\right)rd\varphi dr=\nonumber \\ 
- \left[\int  \Sigma (j - j_\mathrm{p}) (\Omega -\op) rdr \right]^{\Gamma_2}_{\Gamma_1} .
\end{eqnarray}
Here $j=rv_{\varphi}$ is the specific angular momentum and $j_\mathrm{p}$ is $j$ evaluated at the orbital radius
of the protoplanet.

We now follow \cite{1991LPI....22.1463W} and assume the streamline pattern 
is symmetric on the leading and trailing sides such that a 
streamline entering on $\Gamma_2$ can be identified with a corresponding streamline leaving on  $\Gamma_1$
at the same radial location.
  But note that on account of the horseshoe turns, 
these streamlines will  enter ${\cal R}$ at different radii.
 Since, in the barotropic case, potential vorticity or vortensity $\Sigma/\omega$  is conserved along streamlines, 
 and these  streamlines are disconnected, this will  differ on them. Accordingly we write the torque as

\be  \Gamma_\mathrm{c,hs}  =  
 - \int _{\Gamma_2} \Delta 
 \left({\Sigma\over \omega}\right)\omega (j  - j_\mathrm{p}) (\Omega-\op) rdr .\label{HSHOED}\ee
Here 

\be \Delta\left({\Sigma\over \omega}\right) =
 2\left({\Sigma\over \omega}-{\Sigma_\mathrm{p}\over \omega_\mathrm{p}}\right)\ee 
is the vortensity difference on the corresponding streamlines which have been assumed 
to enter ${\cal R}$ at the same radial distance from the planet on opposite sides,
$\omega_\mathrm{p} =\op/2$  is the vorticity at the planets orbital radius and we have assumed
 the vortensity to be an even function of  radial distance from the protoplanet.

The above integral is easily done if one adopts a first order Taylor expansion
for the quantities in brackets and integrates from $\rp-\xs \rp$ to $\rp+\xs \rp$
where
  the dimensionless width of the horseshoe region is  $\xs.$ 
One obtains
\citep{1991LPI....22.1463W}:
\begin{equation}
\label{eqThs}
\Gamma_\mathrm{c,hs}=\frac{3}{4}\left(\frac{3}{2}-\alpha\right)\xs^4 \Sigma_\mathrm{p}\rp^4\op^2.
\end{equation}
Note that as is apparent from the above discussion and that given in the Appendix
and also the results of numerical simulations to be presented later,
 the horseshoe drag, given by  equation (\ref{eqThs}), occurs
 as a non linear effect that has no counterpart in linear theory
 \citep[see also][in addition to \citealt{horse}]{2008A&A...485..877P}.
We also note that  numerical hydrodynamical calculations necessarily have $b>0$, and, 
 two-dimensional simulations,   generally adopt $b\approx h$ to account for three-dimensional 
effects in an approximate way. 
We will see that this strongly affects the width of the horseshoe region $\xs$.
A detailed analysis of the horseshoe region is presented in
 \cite{horse}. Here, we adopt a simple estimate for $\xs$ that has  proved to be reasonable for smoothing lengths comparable to $\hp$ \citep{2008A&A...485..877P}:
\begin{equation}
\xs=\sqrt{\frac{2q}{3b}}.
\label{eqxs}
\end{equation}
In general, this dependence  is to be expected when the separatrix streamline passes through the location of the planet
with $b$ possibly being replaced by $\hp$ for small softening \citep[see also][]{2006ApJ...652..730M}. We will compare the horseshoe drag resulting from $\xs$ to the linear corotation torque in Section \ref{secNonlin}. Note that, since $\Gamma_\mathrm{c,hs}$ is proportional to $\xs^4$, the horseshoe drag  
would then be
proportional to $q^2/\hp^2$ in the small softening case and $q^2/b^2$ in the large softening case
, just as $\Gamma_\mathrm{c,lin}$. This means that the dependence of the total torque on $q,$ $b$ and $\hp$ would not be a good indication of linearity, since both the linear and the non-linear corotation torque scale in the same way. The only ways to distinguish these are through their magnitudes, and by their time evolution.

The linear corotation torque is set up in approximately an orbital time scale
(see the Appendix), similar to the wave torque. This means that, when the background state of the disc is stationary, any evolution in the torque after a few dynamical time scales is due to non-linear effects. The finite width of the horseshoe region gives rise to a \emph{libration} time scale
\begin{equation}
\tau_\mathrm{lib}=\frac{8\pi}{3\xs}\op^{-1},
\end{equation}
which is basically the time it takes for a fluid element at orbital radius $\rp(1+\xs)$ to complete two orbits in a frame corotating with the planet. Therefore, this is the time scale on which the corotation torque will saturate due to phase mixing \citep[see][]{2007LPI....38.2289W}, unless some form of viscosity operates on smaller time scales \citep{2002A&A...387..605M}. Note that saturation is a non-linear process \citep{2003ApJ...587..398O}, since it hinges on the finite width of the corotation region. 

We will see that there are basically three time scales in the problem, two of which are closely related to non-linearity. 
First, there is the orbital time scale, on which all linear torques are set up. 
This is the shortest time scale in the problem. 
The longest time scale is $\tau_\mathrm{lib}$, on which saturation operates. 
A third time scale governs the development of the horseshoe drag, which we will see takes 
a fraction of a libration time \citep[see also][]{2008A&A...485..877P}, 
and lies in between the time scale for the linear torque development and the saturation time scale.   
       

\section{Linear calculations}
\label{secLin}
Since we are interested in departures from linear theory, it is necessary to first firmly establish what linear theory predicts. The 2D linear calculations of  KP93 and TTW02 use a vanishingly small value for $b$, and these results are therefore not directly comparable to our hydrodynamical simulations. Our customised linear calculations, as briefly outlined below, with $b$ comparable to $h$, can be directly compared to hydrodynamical simulations.

\subsection{Governing equations}
Linearising equations \eqref{eqcont} and \eqref{eqmot}, with $\voc{f}_\mathrm{visc}=0$, 
and assuming a Fourier decomposition such that  perturbation quantities  are the real part of a
 sum of terms
 $\propto \exp (im\varphi-im\op t)$, $ m =0,1,2....,$  one obtains the following system of equations
 \citep{1979ApJ...233..857G} for the corresponding Fourier coefficients,
\begin{eqnarray}
\label{eqDum}
i\sgb v'_m - 2\Omega u'_m + \frac{dW'_m}{dr} + \frac{d\Phi'_m}{dr}+\frac{\beta W'_m}{r}=0,\\
\label{eqDum1}i\sgb u'_m+2Bv'_m+\frac{imW'_m}{r}+\frac{im\Phi'_m}{r}=0,\\
\frac{i\sgb W'_m}{\cs^2}+\frac{dv'_m}{dr}+(1-\alpha)\frac{v'_m}{r}+\frac{im u'_m}{r}=0,
\end{eqnarray}
where primes denote perturbation quantities, and the subscript $m,$ being the azimuthal mode number,
 indicates 
the $m$-th Fourier coefficient.
Here the velocity perturbation is ${\voc v}'_ m  =(v'_m, u'_m),$
 $W'_m=\cs^2\Sigma'_m/\Sigma$ is the enthalpy perturbation, $\sgb=m(\Omega-\op)$, $B$ 
is the second Oort constant, and $\Phi'_m$ is the $m$-th Fourier component
 of $\Phi_\mathrm{p}$,   being a real quantity. 
The term proportional to $\beta$ is not present in the equations of KP93,  
because they considered a strictly barotropic equation of state. 
Eliminating $u_m$, we obtain a system of ordinary differential equations (KP93):
\begin{eqnarray}
\frac{dv'_m}{dr}=-\left(1-\alpha-\frac{2mB}{\sgb}\right)\frac{v'_m}{r}+\nonumber \\
i\left(\frac{m^2}{\sgb r^2}-\frac{\sgb}{\cs^2}\right)W'_m+ 
\frac{im^2\Phi'_m}{\sgb r^2} \label{eqlinv} \\
\frac{dW'_m}{dr}=-i\left(\frac{\sgb^2-\kappa^2}{\sgb}\right) v'_m-\nonumber \\
\frac{2m\Omega (W'_m+\Phi'_m)}{\sgb r}-\frac{d\Phi'_m}{dr}-\frac{\beta W'_m}{r},\label{eqlinw}
\end{eqnarray}
where $\kappa^2=4B\Omega$ is the square of the epicyclic frequency.

\begin{figure}
\centering
\resizebox{\hsize}{!}{\includegraphics[]{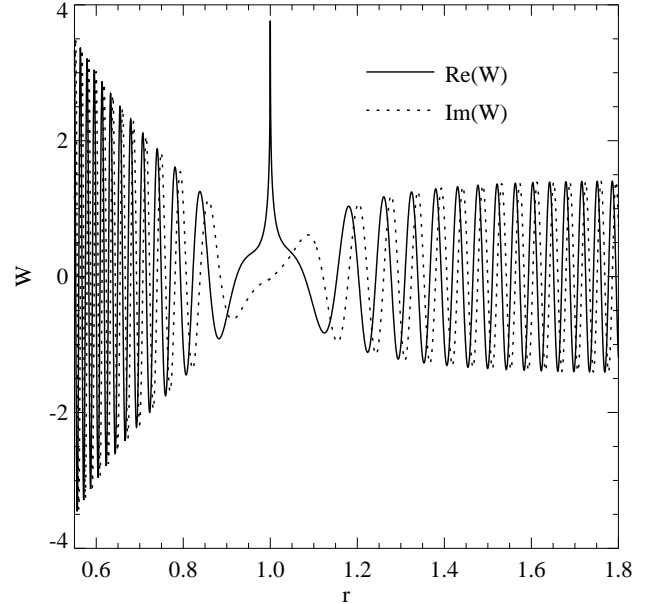}}
\caption{Real (solid) and imaginary (dotted) part of the enthalpy perturbation $W$, for an isothermal, constant surface density disc with $\hp=10^{-3/2}$ and a softening parameter as used in KP93 ($b=10^{-4}$) plotted as a function of radius in units of $\rp.$
}
\label{figenthalpym10}
\end{figure}

We have solved equations \eqref{eqlinv} and \eqref{eqlinw} using a sixth order Runge-Kutta method and outgoing wave boundary conditions (see KP93). In Fig. \ref{figenthalpym10}, we show the resulting $W'_m$ for $m=10$, for an isothermal, constant surface density disc with $\hp=10^{-3/2}$ and a very low value of the softening parameter $b=10^{-4}$. The same case was shown in KP93 (their figure 2), and the results agree very well. 

The imaginary part of $W'_m$ is directly related to the torque density:
\begin{equation}
\frac{d\Gamma_m}{dr}=\pi m \Phi'_m \Sigma \frac{\im(W'_m)}{\cs^2},
\end{equation}
and the total torque on the planet $\Gamma_m$ can be found by integrating over the whole disc. \cite{1979ApJ...233..857G} showed that the Lindblad torque is carried away by density waves, resulting in an angular momentum flux
\begin{eqnarray}
F_m=-\frac{m\pi r\Sigma}{\sgb^2-\kappa^2}\left[\im(W'_m)\frac{d}{dr}(\re(W'_m)+\Phi'_m)-\right. \nonumber\\
\left. (\re(W'_m)+\Phi'_m)\frac{d\im(W'_m)}{dr}\right].
\end{eqnarray}
Therefore, the Lindblad torque is given by $\Delta F_m=F_m(r_\mathrm{out})-F_m(r_\mathrm{in})$, where $r_\mathrm{in}$ and $r_\mathrm{out}$ denote the inner and the outer radius of the disc, respectively. We can therefore calculate the corotation torque as (KP93):
\begin{equation}
\Gamma_{\mathrm{c},m}=\Gamma_m-\Delta F_m,
\end{equation}
and the total torques can be found by summation over all $m$. 

\begin{figure}
\centering
\resizebox{\hsize}{!}{\includegraphics[]{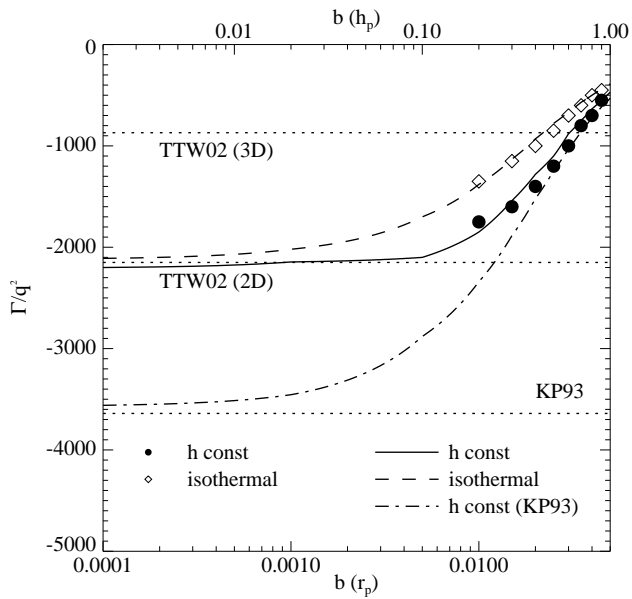}}
\caption{Total torque on the planet, obtained from linear calculations, as a function of the softening parameter $b$. The surface density is proportional to $r^{-3/2}$ ($\alpha=3/2$), which means that the corotation torque vanishes in the barotropic case. The disc is either fully isothermal with $\hp=0.05$ (dashed line) or has a constant aspect ratio $h=0.05$ (solid line). Calculations similar to those of KP93 with constant $h$ are indicated by the dash-dotted line. Also shown are the results of linear calculations in 2D by KP93 (for constant $h$) and the isothermal results of TTW02, for 2D as well as 3D. Both 2D studies used a very small (KP93) or vanishing (TTW02) softening length. Filled circles denote the torques measured from hydrodynamical simulations for a $q=1.26\cdot 10^{-5}$ planet in a disc with constant $h$, and diamonds the results for isothermal simulations with $\hp=0.05$.}
\label{figtorqlin1}
\end{figure}

\subsection{Results}
We start by considering a disc with constant specific vorticity, which means that the corotation torque vanishes. This allows us to look in some more detail at the Lindblad torque alone. In Fig. \ref{figtorqlin1}, we show the total torque\footnote{In all figures, the torque is given in units of $\Sigma_\mathrm{p}\rp^4\op^2$, and is divided by $q^2$ to make it independent of the mass of the planet.} for three different cases: strictly isothermal (dashed line), locally isothermal ($h$ constant; solid line) and locally isothermal without the term proportional to $\beta$ in equation \eqref{eqDum} (dash-dotted line). The latter case is similar to the one considered in KP93, and approaches the result of KP93 for small softening. Similarly, the strictly isothermal case approaches the 2D result of TTW02 for small softening. For appropriate values of $b\approx h$, differences between the strictly and locally isothermal models can be as large as $30\%$. In the remainder of this paper, we will restrict ourselves to the strictly isothermal equation of state, since the horseshoe drag has been analysed for barotropic fluids only \citep{1991LPI....22.1463W}\footnote{In \cite{2008A&A...485..877P}, horseshoe drag was studied in adiabatic flows, where entropy (but not vortensity) is conserved. For a locally isothermal equation of state, neither of these is conserved, with the consequence that it is unclear what the horseshoe drag is in this case.}. For this case, a smoothing length of $b=0.025$, corresponding to $0.5~h$, results in the total torque being equal to the 3D calculations of TTW02.  

\begin{figure}
\centering
\resizebox{\hsize}{!}{\includegraphics[]{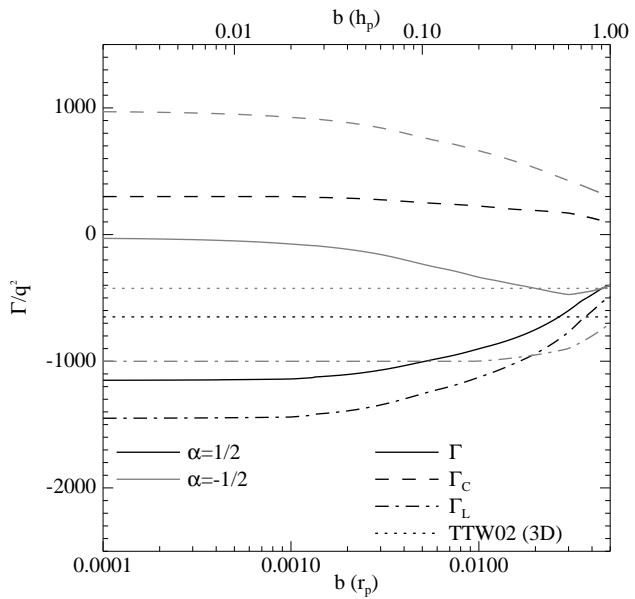}}
\caption{Lindblad (dash-dotted curves) and corotation (dashed curves) torques obtained from an isothermal ($\hp=0.05$) linear calculation as a function of softening parameter $b$ for two different density profiles. Black curves: $\alpha=1/2$, grey curves: $\alpha=-1/2$. The solid curves denote the total torques, and the horizontal dotted lines the 3D results of TTW02.}
\label{figtorqlin2}
\end{figure}

We now introduce a corotation torque in the calculations by considering different values of $\alpha$.
 In Fig. \ref{figtorqlin2}, we show the linear Lindblad and corotation torques for $\alpha=1/2$ and $\alpha=-1/2$.
 Note that in the latter case, the surface density gradient is positive, which may be unrealistic except for special locations
 in the disc \citep[see for example][]{2006ApJ...642..478M}, but it serves as a good example
 of a case with strong corotation torques. For $\alpha=1/2$, the corotation torque is much smaller
 than the Lindblad torque, and is increased by a factor of 3 going from $b=h$ to $b=0$, similar to the Lindblad torque.
 Therefore, the corotation torque is a small fraction of the Lindblad torque for all values of $b$
 considered here. That is not the case for $\alpha=-1/2$, where for small softening parameters the corotation torque
 is almost the same as the Lindblad torque. The sign of the total torque will change for $\alpha<-1/2$, which is expected from the 2D results of KP93 and TTW02.
 For 3D calculations, the situation is different (TTW02), which is illustrated by the results in
 Fig. \ref{figtorqlin2} for larger softening.
 For all values of $\alpha$ considered in Figs. \ref{figtorqlin1} and \ref{figtorqlin2}
 we find that 
 for a softening length of $b=0.5~h$ we can match our 2D
 calculations with the 3D work of TTW02. 

\begin{figure}
\centering
\resizebox{\hsize}{!}{\includegraphics[]{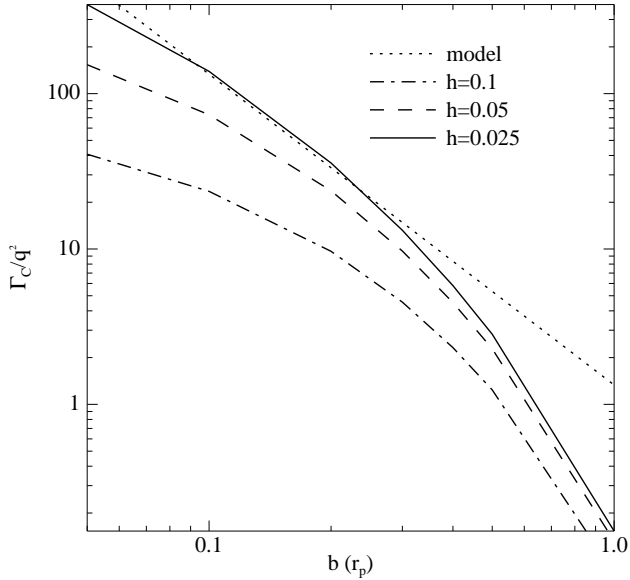}}
\caption{Corotation torque, obtained from a linear calculation with
  $\alpha=1/2$ and $\hp=0.025$ (solid line), $\hp=0.05$ (dashed line),
  and $\hp=0.1$ (dash-dotted line). The dotted line gives the
  prediction of equation \ref{CTsoft}, valid for $h \ll b \ll 1$.} 
\label{figlargesoft}
\end{figure}
\subsection{The limit of large softening}
\label{Totcoro}

We now relate our numerical results to
 the predictions made using the  torque formula (\ref{eqTmlin}) given by
\citet{1979ApJ...233..857G}. This is  applicable to the situation 
where the softening parameter is significantly larger than the scale height
and we  use it to evaluate the corotation torque in that limit.
 To do this we adopt the approximation for 
$\Phi'_m$ introduced by
\citet[]{1980ApJ...241..425G} for $m > 0$ in the form
\be \Phi'_m = {1\over \pi}
\int_0^{2\pi}\Phi_\mathrm{p}\cos(m\varphi)d\varphi  \sim -{2\over \pi}
{G\mpl\over \rp}K_0(m\zeta),\label{Bessel}\ee
where $K_0$ is the standard Bessel function and 

\be \zeta = \sqrt{\frac{(r-\rp)^2}{\rp^2}+b^2}.\ee

\noindent This is valid in the important domain of interest where $m \sim 1/b,$ and
$|\rp-r| < \sim b\rp.$ Summing  the torques given by (\ref{eqTmlin}) over
$m,$ using (\ref{Bessel})  we obtain  
\begin{equation}
\label{eqTmlinsum}
\sum_{m=1}^{\infty} \Gamma_{\mathrm{c},m} = \left({3\over 2}-\alpha\right)
\frac{8 q^2}{3} \Sigma_\mathrm{p}\rp^4 \op^2  \sum_{m=1}^{\infty}m K_0(mb)^2,
\end{equation}
where $ \Sigma_\mathrm{p}$ denotes the unperturbed surface density at the planets location.
Noting that $b$ is small and that the dominant contribution
to the sum on the right hand side comes from large $m \sim 1/b,$
we replace the sum by an integral, thus
 \be \sum_{m=1}^{\infty}m K_0(mb)^2\rightarrow 
\left({1\over b}\right)^2\int^{\infty}_0x K_0(x)^2dx=
{1\over 2} \left({1\over b}\right)^2 .\ee
Thus we obtain
\be \sum_{m=1}^{\infty} \Gamma_{\mathrm{c},m} = \Gamma_\mathrm{c,lin} =
\left({3\over 2}-\alpha\right)
\frac{4 q^2}{3b^2} \Sigma_\mathrm{p}\rp^4 \op^2.\label{CTsoft}.\ee
We remark that \cite{1992NYASA.675..314W} obtained a corresponding expression expression for a vertically averaged potential, that has the same scaling when $b$ is replaced by $\hp$ in the above equation.

A comparison of the torques given by equations (\ref{eqTlin}) and
 (\ref{CTsoft}) indicates that the effect of softening should
 be significant for $b >\sim 1.4 \hp.$
But note that in addition to requiring $b/h \gg 1$ (\ref{CTsoft}) 
also formally requires $b \ll 1.$ This means, as we shall see below,
 that the limit where
(\ref{CTsoft}) applies requires rather small $h < \sim 0.025.$

We now directly compare linear calculations with the prediction of
equation (\ref{CTsoft}). The results are illustrated in
Fig. \ref{figlargesoft}, for three different values of $\hp$. Note
that in the derivation of equation (\ref{CTsoft}) it was assumed that $b
\ll 1$, so we expect the torque to be given by equation (\ref{CTsoft})
for $h\ll b\ll 1$. From Fig. \ref{figlargesoft}, we see that for
$\hp=0.025$ we can nicely reproduce equation (\ref{CTsoft})  for 
 $0.05 <b < 0.2.$
However, as  $\hp$ increases  there are increasing deviations
from  equation (\ref{CTsoft}). For $\hp=0.05$ and  $0.1 < b < 0.2,$
the maximum amd minimum deviations are by a factor of $2$ and $1.5$
respectively, while for $\hp=0.1$ the deviation always exceeds a factor of two.

\begin{figure}
\centering
\resizebox{\hsize}{!}{\includegraphics[]{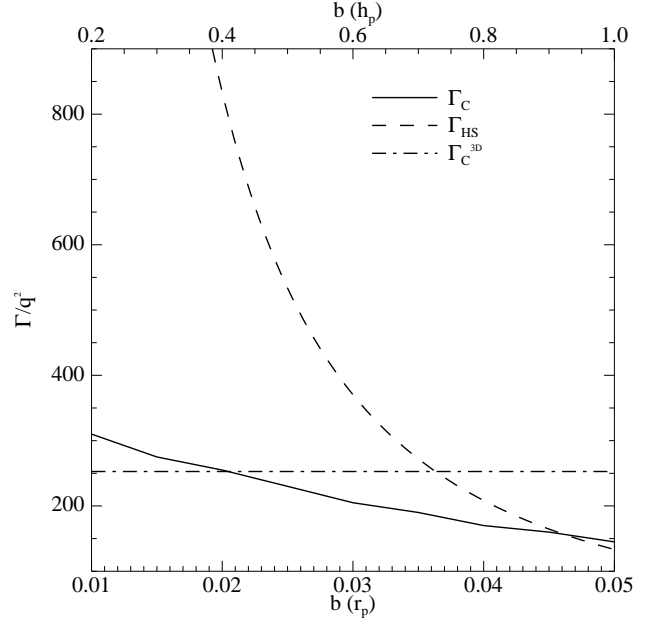}}
\caption{Corotation torque, obtained from a linear calculation with $\alpha=1/2$ and $\hp=0.05$ (solid line), together with the horseshoe orbit drag (dashed line) for our simple estimate of the width of the horseshoe region. Also shown is the 3D result of TTW02 (dash-dotted line).}
\label{figtorqlin3}
\end{figure}


\section{ A comparison of the expected horseshoe drag   with the linear corotation torque}
\label{secNonlin}
   In this section we compare corotation torques obtained from
  linear calculations to  the  expected horseshoe drag.
We  obtain the latter from equation (\ref{eqThs}) which  requires  an estimate of
 $\xs$ which we obtain from  equation (\ref{eqxs}) which simulations have indicated
gives a good estimate for $ b \sim \hp.$ We shall obtain estimates for the horseshoe drag
directly from simulations and make further comparisons below.

In Fig. \ref{figtorqlin3}, we show the linear corotation torque (solid line), 
together with the horseshoe drag, obtained as indicated above, (dashed line) for a disc with $\alpha=1/2$, 
for different values of the smoothing parameter $b$. 
For reasonable values of $b$, the horseshoe drag is stronger than the linear corotation torque 
making the torque on the disc  more negative.   Thus the  horseshoe drag
 on the planet is  positive and  larger than the linear corotation torque. 
For smoothing parameters $0.02 < b < 0.03$, this non-linear torque is a factor $2-3$
 larger than the linear torque. For smaller values of $b$, equation (\ref{eqxs}) predicts a value of $\xs$
that is too large and accordingly  a horseshoe drag that is too large. In reality we expect both the
horseshoe width and therefore the horseshoe drag to reach  limiting values for small $b$
and these values should be larger than those we obtain here  using values of $b$ for
which equation (\ref{eqxs}) applies.
A more complete discussion of these aspects is given in \cite{horse}. 
 Also shown in Fig. \ref{figtorqlin3} is the corotation torque 
 obtained for the corresponding three-dimensional disc (TTW02) which has zero softening. 
 For $b=0.02$, we can match our 2D linear calculation to the 3D result of TTW02. 
 However, for the same softening the horseshoe drag is three times as large.
 
 \noindent For a large softening  with $b=0.7\hp =0.035,$ the horseshoe drag is equal to the 3D 
 unsoftened linear corotation torque. 
 However, numerical results show that for this smoothing length, equation (\ref{eqxs})
 actually slightly  underestimates the true value of $\xs,$   with the consequence that 
 the intersection of the horseshoe drag with the 3D linear corotation torque occurs for
 approximately $b=0.8\hp$. 
 Lacking a full 3D model of the horseshoe region, it is difficult to say what value of $b$ 
 to choose  so that 2D  results   match 3D results. 
 However, numerical results suggest that $b<0.02$ may be appropriate \citep{2006ApJ...652..730M}. 
 We come back to this issue in Section \ref{secDisc}, 
 but it is good to keep in mind that the effects discussed in the next sections 
 may well be stronger in 3D calculations. 


\section{Hydrodynamical simulations}
\label{secNum}
In the previous section, we have established that horseshoe drag is potentially much stronger than the 
linear corotation torque. We now turn to numerical hydrodynamic simulations to show that indeed strong deviations
 from linear theory are encountered in practice. We use the RODEO method \citep{2006A&A...450.1203P} in two spatial dimensions, 
on a regular grid extending from $r=0.5\rp$ to $r=1.8\rp$ and which covers the whole $2\pi$ in azimuth.
 Since we want to resolve the horseshoe region for even the smallest planets we consider, a relatively high
 resolution is required: $1024$ cells in the radial and $4096$ cells in the azimuthal direction,
 making the resolution at the location of the planet approximately $0.0015\rp$ in both directions.
 Tests have shown that this resolution is sufficient to capture the horseshoe dynamics for
 $\xs>0.004$. We include all disc material in the torque calculation
 (not excluding any material close to the planet), and the planet is introduced with its full mass at $t=0$.
 For low mass planets, this does not affect the results.

First, in Section \ref{secDevel}, we study the development of the horseshoe drag and its relationship  to linear theory.
 Then, in Section \ref{secSat}, we study the long-term behaviour (saturation) of the corotation torque. 

\subsection{Development of the non-linear torque} 
\label{secDevel}
We start by considering a planet with a relatively high mass, $q=1.26\cdot 10^{-5}$,
 which corresponds to $4$ $\me$ around a Solar mass star.
 Although this is the  planet with the largest mass we  consider, it  has been expected to be well within the linear regime
 \citep{2006ApJ...652..730M}. 
In Fig. \ref{figtorqlin1}, we show that for a disc with constant specific vorticity,
 we can reproduce the expected linear torque  for all reasonable values of $b$.
 For very small values of $b$, one may expect a departure from linear theory,
 since at this point, an envelope may form that is gravitationally bound to the planet,
 which probably should be excluded from the torque calculation.
 We do not consider this regime here, since we  expect to  adopt  a  value of $b$ comparable to $h$ in order
 to account for vertical averaging. The important point is that we can match our
 linear calculations to the hydrodynamical simulations in the absence of corotation torques
 (linear as well as non-linear).

\begin{figure}
\centering
\resizebox{\hsize}{!}{\includegraphics[]{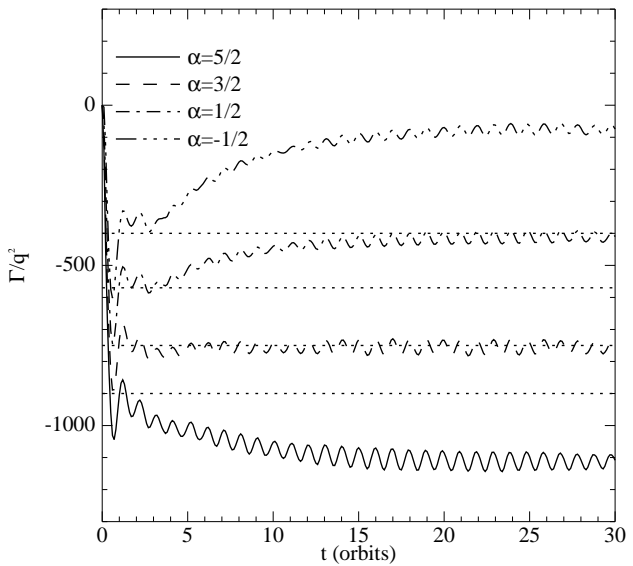}}
\caption{Total torque on the planet for an isothermal, inviscid $\hp=0.05$ disc with $b=0.03$ and $q=1.26\cdot 10^{-5}$, for different surface density profiles. The dotted lines indicate the linear torque, for increasing $\alpha$ from top to bottom.}
\label{figtorqtimalpha}
\end{figure}

This is further illustrated in Fig. \ref{figtorqtimalpha}, where we show the time evolution of the total 
torque on the planet for different surface density profiles. 
We see that the case with $\alpha=3/2$ nicely falls on the corresponding linear result.
 Note also that this torque is set-up in approximately one orbital period.
 This is to be expected for both the Lindblad and the linear corotation torque (see the Appendix).

Different  surface density profiles give remarkably different results. 
All cases except $\alpha=3/2$ show a departure from linear theory, 
the sign of which is dictated by the vortensity gradient.
 This indicates that the corotation torque is enhanced with respect to its linear value.
 This enhancement takes approximately $20$ orbits to develop after which the torques attain steady values
for the remainder of the simulations.
   This time period, as we show below, can be understood as being a fraction of the libration time scale,
 and is therefore related to a non-linear effect.
 Note that in all cases, linear theory is only valid at early times (less than about two orbits)  during which
as expected the linear corotation torque is set up on a dynamical time scale. 
At later times, it gets replaced by the non-linear horseshoe drag.  

\begin{figure}
\centering
\resizebox{\hsize}{!}{\includegraphics[]{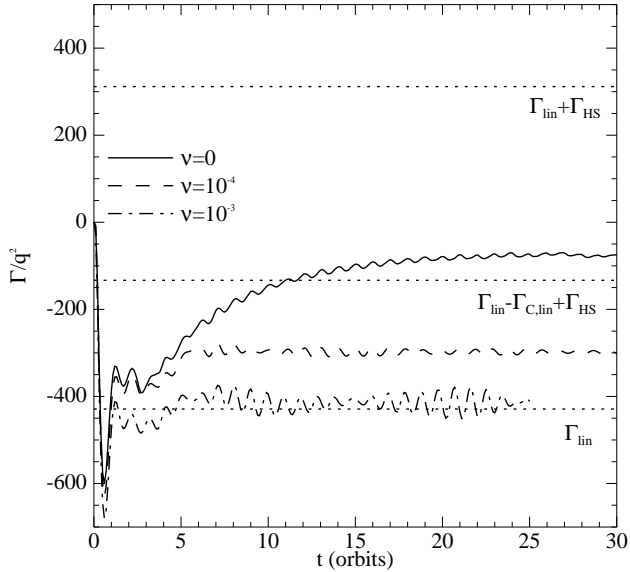}}
\caption{Total torque on the planet in an isothermal $\hp=0.05$ disc with $\alpha=-1/2$ for $b=0.03$ 
and $q=1.26\cdot 10^{-5}$ for different magnitudes of the viscosity.
 The horizontal lines indicate, from bottom to top, the total linear torque, the Lindblad torque with the horseshoe torque 
 (found using the procedure outlined  in Section \ref{secNonlin}) added, 
and the total linear torque with the horseshoe torque added.}
\label{figtorqtim1}
\end{figure}

\begin{figure}
\centering
\resizebox{\hsize}{!}{\includegraphics[]{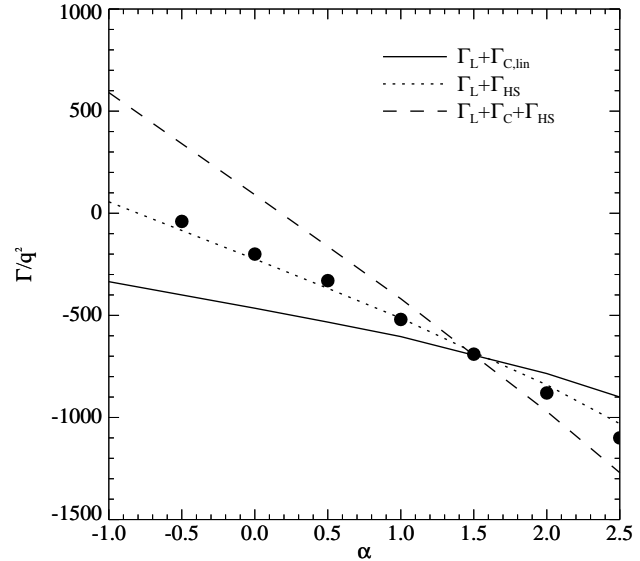}}
\caption{Total torque on a $q=1.26\cdot 10^{-5}$ planet embedded in an
  isothermal $\hp=0.05$ inviscid  disc for $b=0.03$ after $30$
  orbits. The solid line indicates the linear torque, and the dashed
  line shows the total linear torque plus the  estimated horseshoe
  drag. The dotted line indicates the linear Lindblad torque plus the
  non-linear horseshoe drag. Symbols denote results obtained from
  hydrodynamical simulations for different values of the surface
  density power law index $\alpha.$} 
\label{figtorqalphab03iso}
\end{figure}

 We take a closer look at the $\alpha=-1/2$ case illustrated in Fig. \ref{figtorqtim1}. 
 The horizontal dotted lines indicate, from bottom to top, the linear torque, 
 the linear Lindblad torque with the horseshoe drag added,
 and the total linear torque with the horseshoe drag added.
 From the solid line, we see that the total torque can be understood as the linear torque, 
 with the linear corotation torque replaced by the horseshoe drag.
 We have found this to be true for all values of $\alpha$ considered (see Fig. \ref{figtorqalphab03iso}). 
 It is also expected from the discussion given in Section  \ref{secCoro}.

 The magnitude of the non-linear torque depends on the magnitude of the viscosity in the disc. 
 Recall that whenever we include viscosity in the model, we do not allow for large scale mass flow with respect to the planet
 (see also Section \ref{secDisc}). It is easy to understand this dependence,
 since the horseshoe drag model hinges on vortensity conservation while the fluid executes a horseshoe turn.
 For strong enough viscosities, we can recover the linear torque, but note that the large
values  $\nu  > \sim 10^{-3}$ required correspond to a very large
 $\alpha$ viscosity parameter of $\alpha_\mathrm{visc}=0.4$.
 This is because only then can the viscosity directly affect the horseshoe turn.
 In Section \ref{secSat}, we will see that lower viscosities are  in fact sufficient to keep the torque unsaturated.

Returning to inviscid discs,  we show in Fig. \ref{figtorqalphab03iso} the total torque on the disc for different 
values of $\alpha$, confirming that the horseshoe drag indeed replaces the linear corotation torque.
 Note that we expect a torque reversal around $\alpha=-1$, while the linear calculations predict this
 would only happen around $\alpha=-3$. Note also that steep density profiles
 result in an acceleration of inward migration with respect to the linear estimate. 

\begin{figure}
\centering
\resizebox{\hsize}{!}{\includegraphics[]{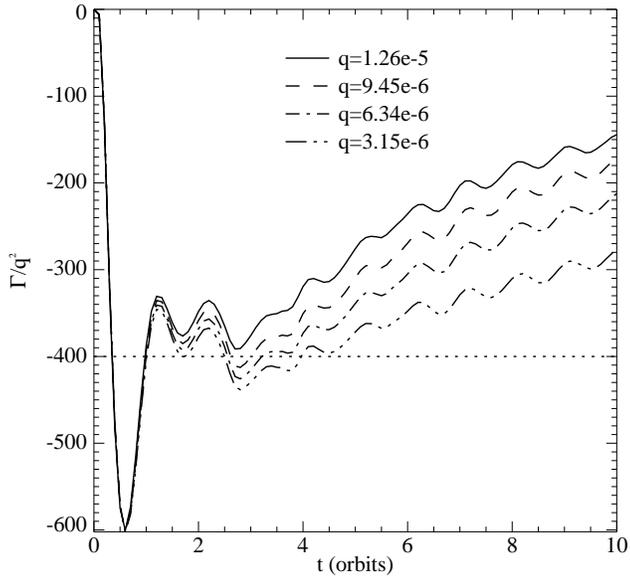}}
\caption{Total torque on a planet in an isothermal $\hp=0.05$ disc for $b=0.03$ and $\alpha=-1/2$, for different values of $q$. The dotted line indicates the linear torque.}
\label{figtorqdiffq}
\end{figure}

Linear theory predicts that the torque should scale as $q^2$. Since in our simple model, 
the horseshoe width scales as $q^{1/2}$, the non-linear contribution to the torque (the horseshoe drag)  also scales as $q^2$.
 Therefore, it could be misinterpreted as a linear effect. However, the time scale on which the horseshoe drag term is set up depends on the mass of the planet. We will see below that is takes a fixed fraction of a libration time scale. This means that although 
the full non-linear torque scales as $q^2$, this will not be the case at intermediate stages.
 This is illustrated in Fig. \ref{figtorqdiffq}, where we show the time evolution of the torque
  for different mass ratios. For linear torques, all curves would fall on top of each other.
 This is indeed true at early times, when the horseshoe drag has not yet developed. 
When the horseshoe drag takes over, however, planets of different mass give different results; 
lower-mass planets take longer to develop the non-linear torque.

\begin{figure}
\centering
\resizebox{\hsize}{!}{\includegraphics[]{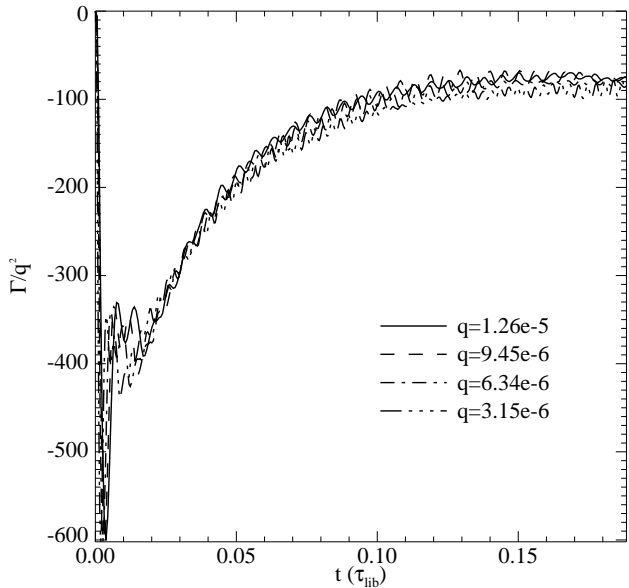}}
\caption{Total torque on the planet embedded in an isothermal $\hp=0.05$ disc for $b=0.03$ and $\alpha=-1/2$, for different values of $q$. The time is in units of $\tau_\mathrm{lib}$, which scales as $q^{-1/2}$.}
\label{figtorqdiffqlib}
\end{figure}

If we rescale the time axis to the libration time (see Fig. \ref{figtorqdiffqlib}), the curves fall on top of each other again.
 This is because the time scale to set up the horseshoe drag is a fixed fraction of the libration time
 (approximately $15\%$, according to Fig. \ref{figtorqdiffqlib}).
  One might expect that, as the scale of the region contributing the torque is of order $H,$
  the time required would be on the order of a factor $\hp$ smaller than the libration time.
  Thus the results presented here are consistent with this time being about $2\hp \tau_{\mathrm{lib}}.$ 
 This  type of phenomenon was also discussed in \cite{2008A&A...485..877P}, where it was shown that the development of the non linear torque is due to an high density ridge resulting from material that has executed a horseshoe turn at constant entropy (replacing  specific vorticity applicable to this case). 

\begin{figure}
\centering
\resizebox{\hsize}{!}{\includegraphics[]{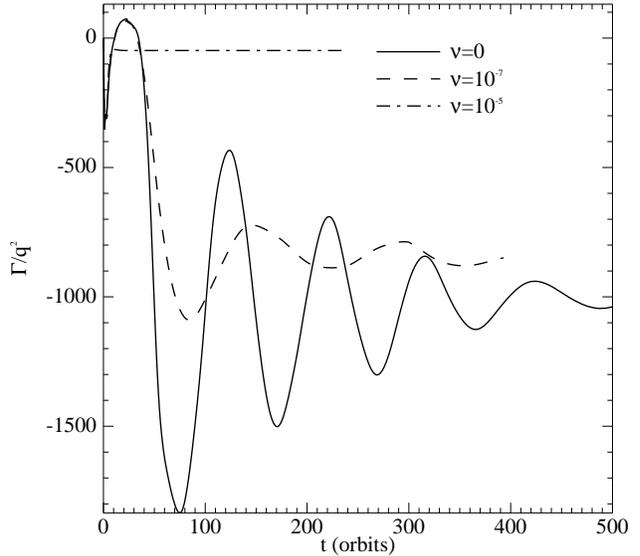}}
\caption{Total torque on a planet in an isothermal $\hp=0.05$ disc for $q=1.26\cdot 10^{-5}$, $b=0.02$ and $\alpha=-1/2$, 
for different values of the viscosity $\nu$.}
\label{figtorqsat}
\end{figure}
\begin{figure}
\centering
\resizebox{\hsize}{!}{\includegraphics[]{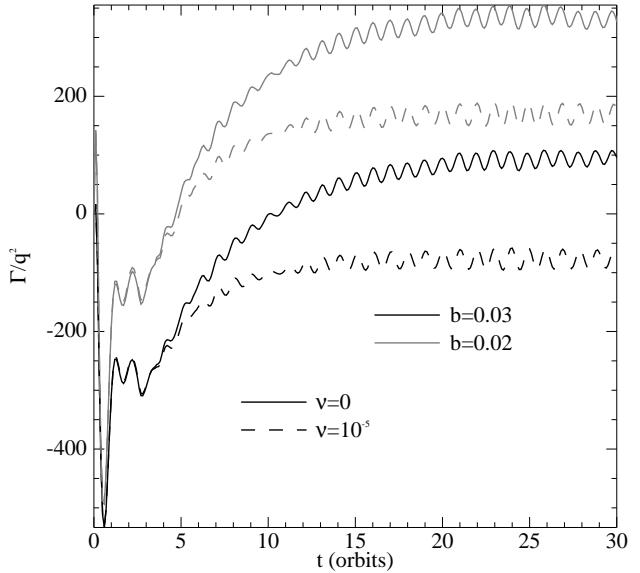}}
\caption{Total torque on a planet in an isothermal $\hp=0.05$ disc for $q=1.26\cdot 10^{-5}$ and $\alpha=-1$, for different values of softening parameter $b$ and viscosity $\nu$.}
\label{figtorqtimb}
\end{figure}

\subsection{Long-term evolution}
\label{secSat}
Lindblad torques give rise to waves that carry away angular momentum. 
Therefore, the planet can continue to exchange angular momentum with the disc at Lindblad resonances. 
However, there  is no wave  transport in the corotation region, 
and therefore  in the absence of any other form of transport,
 only a finite amount of angular momentum exchange with the planet can occur before
the structure of the coorbital region is significantly modified. 
 In other words, the corotation region is a closed system unless there is some diffusive process operating in the disc
that can  transport angular momentum there. 
In the absence of such  diffusion, the corotation torque will saturate \citep{2003ApJ...587..398O}. 

Saturation is an inherently non-linear process, as it depends again on a finite width of the corotation region. 
The time scale on which saturation operates is the libration time scale, 
and diffusion must operate on a smaller time scale in order to prevent saturation. For a viscosity coefficient $\nu$, one needs
\begin{equation}
\nu > \frac{\xs^3}{4} \rp^2\op,
\end{equation} 
\citep[see][who considered this process for a disc with constant surface density]{2002A&A...387..605M}. 
For a level of viscosity that has become standard in
 disc-planet interaction studies, $\nu=10^{-5}$ $\rp^2\op$,
  all protoplanets considered here are of small enough mass that we expect the corotation torque to be unsaturated.
 We consider a planet of $q=1.26\cdot 10^{-5}$, and use a softening
 parameter $b=0.02$ to make the libration time scale as short as possible to ease the computational burden.
In Fig. \ref{figtorqsat}, we show the long-term evolution of the torque for three different levels of viscosity. 
The inviscid case shows strong libration cycles, before the torque starts to settle to a value that is close to the
linear  Lindblad torque ($\Gamma_\mathrm{L}\approx -1000 q^2$, see Fig. \ref{figtorqlin2}). 

A small viscosity of $\nu=10^{-7}$ gives less prominent libration cycles, and the torque settles at a value of approximately $-850$.
 In this case, corotation torques are partially saturated. This is in quantitative agreement with the analysis of \cite{2001ApJ...558..453M}, where it is argued that the horseshoe drag should be multiplied by a factor $\mathcal{F}(z)$ with $z=\xs (2\pi\nu)^{-1/3}$ to account for saturation.  Choosing the smallest option for $\mathcal{F}$ given in \cite{2001ApJ...558..453M}, which gave the best fit in \cite{2002A&A...387..605M}, we find $\mathcal{F}=0.114$. This predicts a corotation torque of $\Gamma_\mathrm{C}=125$, while the difference between the Lindblad torque and the total torque as measured from the simulations (the asymptotic difference between the dashed and solid curve in Fig. \ref{figtorqsat}) is 150. 
 
For $\nu=10^{-5}$ the libration cycles disappear, indicating that
 the torque remains unsaturated. There is basically no evolution of  the torque once the non linear contribution has been set up. This again agrees with the analysis of \cite{2001ApJ...558..453M}, where for this value of $\nu$ we expect $\mathcal{F}$ close to unity. Note, however, that the maximum torque that can be reached is reduced for this relatively high viscosity (see also Fig. \ref{figtorqtim1}).

Finally, we show in Fig. \ref{figtorqtimb} that, for a softening parameter $b=0.02$,
 low mass planets will feel a sustained positive torque when $\alpha=-1$,
 which corresponds to a mildly positive surface density gradient.
 Although a viscosity $\nu=10^{-5}$ reduces the non-linear torque, making the torque negative
 for $b=0.03$, it is necessary for the torque to be sustained.
 For a slightly smaller softening parameter, $b=0.02$,
 we find sustained outward migration. Note that from Figs. \ref{figtorqlin1}
 and \ref{figtorqlin2} we expect a smoothing around $b=0.025$ to reproduce 3D effects.
 Full 3D simulations, together with a 3D understanding of horseshoe dynamics,
 are needed to see how this non-linear torque behaves in a three-dimensional setting (see also Section \ref{secDisc}).
  
\begin{figure}
\centering
\resizebox{\hsize}{!}{\includegraphics[]{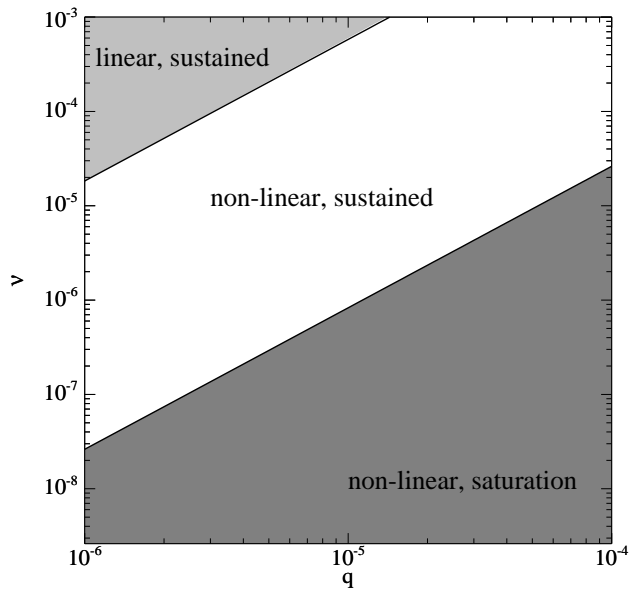}}
\caption{Schematic overview of the different migration regimes. Depending on the magnitude of the viscosity, low-mass planets will either experience a linear, unsaturated torque, a non-linear, unsaturated torque, or a non-linear, saturated torque.}
\label{figqnu}
\end{figure}

\section{Discussion}
\label{secDisc}
We have shown that non-linear effects commonly occur in the coorbital region, even for low-mass planets. In inviscid discs, the corotation torque eventually \emph{always}  becomes non-linear,
  saturating after a few libration cycles, so that only the (linear) Lindblad torque survives.
Viscosity can prevent the torque from saturating, and for a large enough viscosity,
which however decreases with protoplanet mass, the linear torque will be restored.
In Fig. \ref{figqnu} we present a schematic
 overview of the three possibilities in the $(\log q, \log \nu )$ plane.
Two lines define the boundaries between the three regions.
The boundary between the region where the non linear torque is maintained by viscosity 
and the saturated regime is 
given approximately by the condition that viscous diffusion across the coorbital
region occurs in one libration cycle. 
 The line separating  the sustained horseshoe drag regime
from the linear regime  is expected to occur  where viscous effects 
become large enough to disrupt horseshoe turns.

  Note that the  regime for which the corotation torque is linear, occupies only a small fraction of the parameter space. Note also that the borders between the different regions are not razor-sharp. In reality the non-linear torque can be partially saturated, or be reduced by viscosity. However, the overall picture is clear: the corotation torque is almost always non-linear.

For density profiles with  close to $\alpha=3/2$, the non-linear corotation  torque, or horseshoe drag, does not play a major role.
 In an isothermal  disc with $\alpha=1/2$, a  commonly  adopted value for  a locally isothermal disc,
 the deviation from the linear torque can be up to  $30\%$  in the inviscid case (see Fig. \ref{figtorqtimalpha}). 
 If a  relatively large viscosity is used, such non-linear behaviour can be markedly reduced  \citep[see also the discussion in][]
{2008A&A...485..877P}. Note also that for a given viscosity, linearity is always restored for small enough protoplanet masses (see Fig. \ref{figqnu}). Large viscosities,
 numerical and/or imposed, in combination with 
inadequate numerical resolution for representing  horseshoe turns for low-mass planets, 
 conspire to make  the non-linear behaviour described here less apparent.

\cite{2006ApJ...652..730M} reported non-linear behaviour for planets of higher mass than considered here, 
also claimed to be due to the action of the horseshoe drag. 
However, since the horseshoe width $\xs$ was enhanced relative to its value for low-mass planets, the non-linear torque was much stronger. \cite{2006ApJ...652..730M} define non-linearity as a departure from the torque being proportional to $q^2$. In the light of our findings, however, it should be noted that \emph{all} planets in fact show non-linear behaviour in the inviscid limit. For low-mass planets, this non-linearity is less obvious because the horseshoe drag is also proportional to $q^2$.

For moderately positive density gradients ($\alpha \leq -1$), the non-linear torque is strong enough to reverse the sign of the total torque. In realistic discs, positive gradients will probably only be realised at special locations, but note that, unlike as was argued in \cite{2006ApJ...642..478M}, the gradient does not have to be extremely sharp to stop inward migration. Almost any positive surface density gradient can act as a 'protoplanet trap'.

Barotropic discs with $\alpha<0$ can also serve as a model for what happens when strong corotation torques arise that are not due to the radial vortensity gradient. In adiabatic discs, for example, there is a strong contribution from any radial entropy gradient \citep{2008A&A...478..245P,2008A&A...485..877P}. \cite{2008ApJ...672.1054B} showed that linear theory predicts a contribution from a radial entropy gradient, and argued that this would be strong enough to reverse the total torque. However, \cite{2008A&A...485..877P} showed that this linear contribution is in fact small, and that it is a non-linear effect that changes the sign of the torque as observed in \cite{2006A&A...459L..17P}, and more recently in \cite{2008arXiv0806.2990K}. The non-linear contribution arises in a similar manner
to that  discussed in this paper \citep[see also][]{2008A&A...485..877P}.

We have greatly simplified the problem by keeping the planet on a fixed orbit and choosing the viscosity law such that there is no accretion flow. This way, there is no radial mass flow with respect to the planet. Material that flows past the planet from the inner to the outer disc (or the other way round) exerts an additional torque on the planet \citep[see][]{2001ApJ...558..453M}. The effect of such a radial flow is to introduce an asymmetry in the horseshoe region, which lies at the basis of Type III migration \citep{2003ApJ...588..494M}. It remains to be investigated how these processes affect the current analysis. 

\cite{2006ApJ...652..730M} noted that the effects of non-linearity were much stronger in 3D, probably because the width of the horseshoe region is larger compared to 2D simulations that necessarily have a softening parameter of the order of the scale height of the disc. The full 3D structure of the horseshoe region remains to be investigated, but if the velocity field is essentially two-dimensional, one could regard the vertical structure of the horseshoe region as stacked layers of 2D horseshoes, with a smoothing equal to the vertical distance to the midplane \citep[see also][]{2006ApJ...652..730M}. Then, the term $\xs^4$ in equation \eqref{eqThs} would be replaced by a density-weighted average (see \cite{2002A&A...387..605M}):
\begin{equation}
\xs^4 \rightarrow \frac{\int_{-\infty}^\infty \rho(\rp,\varphi_\mathrm{p},z) \xs^4(b=z)dz}{\Sigma_\mathrm{p}}.
\end{equation}
However, this procedure requires a proper estimate for $\xs$ for $b\rightarrow 0$, where equation \eqref{eqxs} cannot be used. This issue is discussed in a   paper. 


\section{Summary and conclusion}
\label{secCon}
We have analysed the corotation torque on an embedded planet in a barotropic disc, and shown that this torque is non-linear in general after a few orbits. The linear corotation torque, which is set up on a dynamical time scale, is replaced by horseshoe drag, which is stronger in all cases we have considered. This process completes in approximately one tenth of the libration time scale $\tau_\mathrm{lib}$
or $\sim 2\hp \tau_{\mathrm{lib }}$ in our case. 

For discs with large vortensity gradients, a strong departure from linear theory is observed. We have shown that in discs with moderately positive density gradients ($\Sigma \propto r$)  non-linear effects can reverse the total torque, leading to outward migration. In particular, this may
occur without a very abrupt strong surface density transition of the type considered by \citep['protoplanetary trap',][]{2006ApJ...642..478M}, and may lead to the
halting of the inward migration of low-mass planets. 

After one libration time, saturation sets in unless some form of viscosity is able to restore the original density profile within $\tau_\mathrm{lib}$. For the standard value of $\nu=10^{-5}$, corresponding to $\alpha_\mathrm{visc}=0.004$, the corotation torque on low-mass planets in the Earth mass
range can be sustained by such action.
  
\section*{Acknowledgements}
We thank W. Kley for useful comments, and the anonymous referee for an insightful report. This work was performed using the Darwin Supercomputer of the University of Cambridge High Performance Computing Service (http://www.hpc.cam.ac.uk), provided by Dell Inc. using Strategic Research Infrastructure Funding from the Higher Education Funding Council for England.

\bibliography{paardekooper.bib}

\begin{thebibliography}{}

\bibitem[\protect\citeauthoryear{{Baruteau} \& {Masset}}{{Baruteau} \&
  {Masset}}{2008}]{2008ApJ...672.1054B}
{Baruteau} C.,  {Masset} F.,  2008, \apj, 672, 1054

\bibitem[\protect\citeauthoryear{{Crida} \& {Morbidelli}}{{Crida} \&
  {Morbidelli}}{2007}]{2007MNRAS.377.1324C}
{Crida} A.,  {Morbidelli} A.,  2007, \mnras, 377, 1324

\bibitem[\protect\citeauthoryear{{D'Angelo}, {Henning} \& {Kley}}{{D'Angelo}
  et~al.}{2002}]{2002A&A...385..647D}
{D'Angelo} G.,  {Henning} T.,    {Kley} W.,  2002, \aap, 385, 647

\bibitem[\protect\citeauthoryear{{Goldreich} \& {Tremaine}}{{Goldreich} \&
  {Tremaine}}{1979}]{1979ApJ...233..857G}
{Goldreich} P.,  {Tremaine} S.,  1979, \apj, 233, 857

\bibitem[\protect\citeauthoryear{{Goldreich} \& {Tremaine}}{{Goldreich} \&
  {Tremaine}}{1980}]{1980ApJ...241..425G}
{Goldreich} P.,  {Tremaine} S.,  1980, \apj, 241, 425

\bibitem[\protect\citeauthoryear{{Kley} \& {Crida}}{{Kley} \&
  {Crida}}{2008}]{2008arXiv0806.2990K}
{Kley} W.,  {Crida} A.,  2008, \aap, 487, L9

\bibitem[\protect\citeauthoryear{{Korycansky} \& {Pollack}}{{Korycansky} \&
  {Pollack}}{1993}]{1993Icar..102..150K}
{Korycansky} D.~G.,  {Pollack} J.~B.,  1993, Icarus, 102, 150

\bibitem[\protect\citeauthoryear{{Lai} \& {Zhang}}{{Lai} \&
  {Zhang}}{2006}]{2006MNRAS.368...917Z}
{Lai} D.,  {Zhang} H.,  2006, \mnras, 368, 917

\bibitem[\protect\citeauthoryear{{Lin} \& {Papaloizou}}{{Lin} \&
  {Papaloizou}}{1986}]{1986ApJ...309..846L}
{Lin} D.~N.~C.,  {Papaloizou} J.,  1986, \apj, 309, 846

\bibitem[\protect\citeauthoryear{{Masset}}{{Masset}}{2001}]{2001ApJ...558..453%
M}
{Masset} F.~S.,  2001, \apj, 558, 453

\bibitem[\protect\citeauthoryear{{Masset}}{{Masset}}{2002}]{2002A&A...387..605%
M}
{Masset} F.~S.,  2002, \aap, 387, 605

\bibitem[\protect\citeauthoryear{{Masset}, {D'Angelo} \& {Kley}}{{Masset}
  et~al.}{2006}]{2006ApJ...652..730M}
{Masset} F.~S.,  {D'Angelo} G.,    {Kley} W.,  2006, \apj, 652, 730

\bibitem[\protect\citeauthoryear{{Masset}, {Morbidelli}, {Crida} \&
  {Ferreira}}{{Masset} et~al.}{2006}]{2006ApJ...642..478M}
{Masset} F.~S.,  {Morbidelli} A.,  {Crida} A.,    {Ferreira} J.,  2006, \apj,
  642, 478

\bibitem[\protect\citeauthoryear{{Masset} \& {Papaloizou}}{{Masset} \&
  {Papaloizou}}{2003}]{2003ApJ...588..494M}
{Masset} F.~S.,  {Papaloizou} J.~C.~B.,  2003, \apj, 588, 494

\bibitem[\protect\citeauthoryear{{Mayor} \& {Queloz}}{{Mayor} \&
  {Queloz}}{1995}]{1995Natur.378..355M}
{Mayor} M.,  {Queloz} D.,  1995, \nat, 378, 355

\bibitem[\protect\citeauthoryear{{Nelson} \& {Papaloizou}}{{Nelson} \&
  {Papaloizou}}{2004}]{2004MNRAS.350...849N}
{Nelson} R.~P.,  {Papaloizou} J.~C.~B.,  2004, \mnras, 350, 849

\bibitem[\protect\citeauthoryear{{Ogilvie} \& {Lubow}}{{Ogilvie} \&
  {Lubow}}{2003}]{2003ApJ...587..398O}
{Ogilvie} G.~I.,  {Lubow} S.~H.,  2003, \apj, 587, 398

\bibitem[\protect\citeauthoryear{{Paardekooper} \& {Mellema}}{{Paardekooper} \&
  {Mellema}}{2006a}]{2006A&A...459L..17P}
{Paardekooper} S.-J.,  {Mellema} G.,  2006a, \aap, 459, L17

\bibitem[\protect\citeauthoryear{{Paardekooper} \& {Mellema}}{{Paardekooper} \&
  {Mellema}}{2006b}]{2006A&A...450.1203P}
{Paardekooper} S.-J.,  {Mellema} G.,  2006b, \aap, 450, 1203

\bibitem[\protect\citeauthoryear{{Paardekooper} \& {Mellema}}{{Paardekooper} \&
  {Mellema}}{2008}]{2008A&A...478..245P}
{Paardekooper} S.-J.,  {Mellema} G.,  2008, \aap, 478, 245

\bibitem[\protect\citeauthoryear{{Paardekooper} \& {Papaloizou}}{{Paardekooper}
  \& {Papaloizou}}{2008}]{2008A&A...485..877P}
{Paardekooper} S.-J.,  {Papaloizou} J.~C.~B.,  2008, \aap, 485, 877

\bibitem[\protect\citeauthoryear{{Paardekooper} \& {Papaloizou}}{{Paardekooper}
  \& {Papaloizou}}{2009}]{horse}
{Paardekooper} S.-J.,  {Papaloizou} J.~C.~B.,  2009, \mnras, submitted

\bibitem[\protect\citeauthoryear{{Papaloizou}, {Nelson}, {Kley}, {Masset} \&
  {Artymowicz}}{{Papaloizou} et~al.}{2007}]{2007prpl.conf..655P}
{Papaloizou} J.~C.~B.,  {Nelson} R.~P.,  {Kley} W.,  {Masset} F.~S.,
  {Artymowicz} P.,  2007, in Protostars and Planets V {Disk-Planet Interactions
  During Planet Formation}.
pp 655--688

\bibitem[\protect\citeauthoryear{{Pepli{\'n}ski}, {Artymowicz} \&
  {Mellema}}{{Pepli{\'n}ski} et~al.}{2008}]{2008MNRAS.387.1063P}
{Pepli{\'n}ski} A.,  {Artymowicz} P.,    {Mellema} G.,  2008, \mnras, 387, 1063

\bibitem[\protect\citeauthoryear{{Tanaka}, {Takeuchi} \& {Ward}}{{Tanaka}
  et~al.}{2002}]{2002ApJ...565.1257T}
{Tanaka} H.,  {Takeuchi} T.,    {Ward} W.~R.,  2002, \apj, 565, 1257

\bibitem[\protect\citeauthoryear{{Terquem}}{{Terquem}}{2003}]{2003MNRAS.341.11%
57T}
{Terquem} C.~E.~J.~M.~L.~J.,  2003, \mnras, 341, 1157

\bibitem[\protect\citeauthoryear{{Ward}}{{Ward}}{1989}]{1989ApJ...336..526W}
{Ward} W.~R.,  1989, \apj, 336, 526

\bibitem[\protect\citeauthoryear{{Ward}}{{Ward}}{1991}]{1991LPI....22.1463W}
{Ward} W.~R.,  1991, in Lunar and Planetary Institute Conference Abstracts
  {Horsehoe Orbit Drag}.
p.~1463

\bibitem[\protect\citeauthoryear{{Ward}}{{Ward}}{1992}]{1992NYASA.675..314W}
{Ward} W.~R.,  1992, in {Dermott} S.~F.,  {Hunter} Jr. J.~H.,   {Wilson} R.~E.,
   eds, Astrophysical Disks Vol.~675 of New York Academy Sciences Annals,
  {Disk-Protoplanet Interactions: Torques from the Coorbital Zone}.
pp 314--+

\bibitem[\protect\citeauthoryear{{Ward}}{{Ward}}{1997}]{1997Icar..126..261W}
{Ward} W.~R.,  1997, Icarus, 126, 261

\bibitem[\protect\citeauthoryear{{Ward}}{{Ward}}{2007}]{2007LPI....38.2289W}
{Ward} W.~R.,  2007, in Lunar and Planetary Institute Conference Abstracts
  Vol.~38 of Lunar and Planetary Institute Conference Abstracts, {A Streamline
  Model of Horseshoe Torque Saturation}.
p.~2289

\end{thebibliography}

\label{lastpage}

\section*{Appendix}
\section*{Time development of the linear corotation torque}
We here consider the time development of the linear corotation
torque consequent on the insertion of a perturbing protoplanet potential
into an unperturbed disc. We specialise to the limit of a low mass
planet and adopt a local Cartesian coordinate system $(x,y)$
with origin  at the centre of mass of the planet, and  the $x$ axis pointing
radially outwards.  The system rotates uniformly
with the Keplerian angular velocity, $\op,$
 at  the orbital location of the planet. In this frame the differential
rotation of the disc is manifest through a linear shear 
$ \voc v = (v_x, v_y) = (0, -3\op x/2).$
Thus the background vorticity is constant and the vortensity gradient is
proportional to the gradient of $1/\Sigma.$

The governing equations are the inviscid form
of the basic equations (\ref{eqcont}) and 
(\ref{eqmot}).
As it builds up from nothing, the disc response has to be 
linear for some period of time
after insertion of the protoplanet so we linearise the  governing equations
treating $\Phi_\mathrm{p}$ as a linear perturbation.
The linear response approaches its final value in a characteristic time
independent of the protoplanet mass, in fact we shall see
that this is characteristically the orbital time scale. Accordingly 
a full linear response is guaranteed for a sufficiently small protoplanet mass.

For simplicity we initially discuss the large softening  case
for which the pressure response is small (see Section \ref{Totcoro}).
We then go on to discuss the more general case.

\subsection*{Calculation of the linear response}
The linearised forms of the components of (\ref{eqmot}) in the local system 
are

\be  { D v_x'\over Dt }
-2\op v_y'
=  - {\partial\Phi_\mathrm{Gp}\over \partial x} \label {motloclinx}, \hspace{0.5cm} {\rm and}\ee

\be   {  D v_y' \over Dt}+
{1\over 2}\op v_x'
=  - {\partial\Phi_\mathrm{Gp} \over \partial y} \label {motlocliny} ,\ee
where perturbation quantities are denoted with a prime and
the convective derivative  applies to  the unperturbed flow
such  that

 \be {D\over D t} \equiv { \partial \over \partial t} -{3\op x\over 2} 
{ \partial \over \partial y}.\ee
These are the local analogues of the global linear equations (\ref{eqDum}) and (\ref{eqDum1}),
$\Phi_\mathrm{Gp}$ here denotes the generalised potential which may be 
taken to be the sum of the perturbing protoplanet potential
and the enthalpy perturbation. 
For the large softening model,
$\Phi_\mathrm{Gp}$ may be taken to be the protoplanet perturbing potential alone.
Note we have not made a Fourier decomposition at this point.

We find it convenient to work with the Lagrangian displacement 
{\mbox{\boldmath$\xi$}} $\equiv$
$(\xi_x,\xi_y)$, which is such that 
$(v_x',v_y') = (D\xi_x/Dt, D\xi_y/Dt+ 3\op\xi_x/2).$
In terms of this (\ref{motloclinx}) and (\ref{motlocliny}) take the form

\be  { D ^2\xi_x\over Dt }
-2\op{ D \xi_y\over Dt } -3\op^2\xi_x 
=  - {\partial\Phi_\mathrm{Gp}\over \partial x} \label {motloclinLx}, \hspace{0.5cm} {\rm and}\ee

\be   {  D^2 \xi_y \over Dt^2}+2\op{ D \xi_x\over Dt }
=  - {\partial\Phi_\mathrm{Gp} \over \partial y} \label {motloclinLy} ,\ee
We now perform a Fourier decomposition of $\Phi_\mathrm{Gp}$ in $y$ assuming periodicity
on a length scale $L_y.$
\be \Phi_\mathrm{Gp} ={\cal R}e\sum_{n=0}^{\infty} b_n(x,t) \exp(ink_{y0}y),\label{Four}\ee
where $k_{y0}=2\pi/L_y$ and ${\cal R}e$  denotes that the real part is to be taken.
We remark that to relate to  a full  cylindrical annulus, $L_y \rightarrow  2\pi \rp,$   
$y \rightarrow \rp\varphi,$ and $n \rightarrow m,$ where $m$ is the usual azimuthal
mode number. Using this we obtain 
\be  { D ^2\xi_{x}\over Dt }
-2\op{ D \xi_{y}\over Dt } -3\op^2\xi_{x}
=  -\sum_{n=0}^{\infty}{\partial  b_{n}\over \partial  x} \exp(ink_{y0}y)
 \label {motlocfLx}, \hspace{0.5cm} {\rm and}\ee
\be   {  D^2 \xi_{y} \over Dt^2}+2\op{ D \xi_{x}\over Dt }
= -\sum_{n=0}^{\infty}ink_{y0}b_{n} \exp(ink_{y0}y)  \label {motlocfLy} .\ee

\begin{figure}
\centering
\resizebox{\hsize}{!}{\includegraphics[]{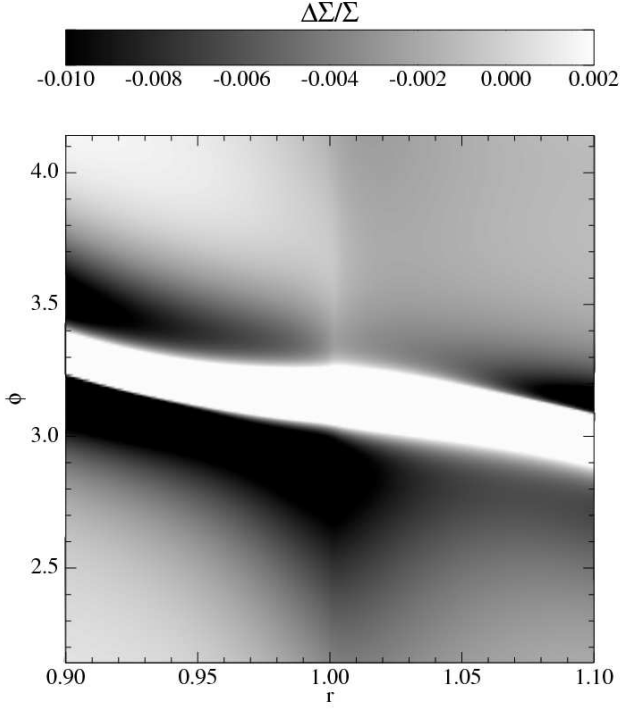}}
\caption{Linear density response for $\alpha=-2$, $b=0.03$, $\hp=0.05$ 
  and $q=1.26\cdot 10^{-5}$. The color scale has been adjusted to
  highlight the density ridge at corotation, which gives rise to the
  linear corotation torque. The planet is located at $r=1$
and $\phi=\pi.$} 
\label{figlindens}
\end{figure}

In order to solve these equations we make the approximation
of neglecting $D^2\xi_x/Dt^2.$  This is a common approximation
that is made when considering corotation/horseshoe dynamics. 
It has the effect of removing   epicyclic   oscillations, and therefore
Lindblad torques,  but as we shall see it
does not interfere with the corotation torque.
We then use (\ref{motlocfLx}) to eliminate $\xi_x$ from (\ref{motlocfLy})
and thus obtain a  second order equation for $\xi_y:$
\be   {1\over 3}{ D^2 \xi_{y} \over Dt^2}
= -\sum_{n=0}^{\infty} c_{n} \exp(ink_{y0}y)
\label{motlocfLy1} ,\ee
where
\be c_n = -{2\over 3\op}{\partial^2 b_{n}\over \partial x \partial t}- ink_{y0}( b_{n} -xdb_{n}/dx).\ee
This may be  integrated to obtain $\xi_y$ and then $\xi_x$ found from the used
approximate form of (\ref{motlocfLx}). The integration  process is aided
by noting that  $y+3\op xt/2$ is invariant under $D.$
Note too  that in order to ensure that  the boundary condition, that
all disturbances vanish at $t=0,$ is satisfied,  an appropriate function of $y+3\op xt/2,$
which plays  the role of an integration constant,
may be added ($D/Dt$ operating on such a function  will be zero).  
Following the  above procedures, which 
guarantee there will be no singularities in the solution, we obtain
$${ D \xi_{y} \over Dt}
=-3\sum_{n=0}^{\infty}\exp(ink_{y0}
(y+3\op xt/2))\times $$  
\be \hspace{3cm}\int^{t}_{0} c_n\exp(-3ink_{y0}\op xt'/2))dt' ,
\label{solocfLy} \ee
\be \xi_{x}=-{2\over 3\op}{ D \xi_{y}\over Dt } 
+{1\over 3\op^2}\sum_{n=0}^{\infty}{d b_{n}\over d x}\exp(ink_{y0}y)\ee
and
$$ \xi_{y}
=-3\sum_{n=0}^{\infty}\exp(ink_{y0}
(y+3\op xt/2))\times $$  
\be \hspace{2.5cm} \int^{t}_{0} c_n\exp(-3ink_{y0}\op xt'/2))(t-t')dt'.\ee
We note that in the above  and other similar integrands, unless otherwise
indicated, quantities are evaluated at the time  $t'$.
We may now find the density perturbation (see
  Fig. \ref{figlindens} for and example) from
\be \Sigma' = -\nabla\cdot\Sigma {\mbox{\boldmath$\xi$}} \ee
and then evaluate the corotation torque acting on the planet,
$\Gamma_\mathrm{c,lin},$  by doing the torque integral
\be  \Gamma_\mathrm{c,lin} = \rp \int {\cal R} e 
(\Sigma'){\cal R}e\left( {\partial \Phi_\mathrm{Gp}\over \partial y}\right) dxdy,\ee
where  we have adopted a multiplicative factor
equal to an orbital radius $\rp$ in order to convert a force in the
$y$ direction into a torque. 
Substituting  
the Fourier expansion of $\partial \Phi_\mathrm{Gp}/\partial y$ and integrating over $y$
we find,  after some elementary algebra,  that $ \Gamma_\mathrm{c,lin} = \Gamma_\mathrm{a} +\Gamma_\mathrm{b},$ where
 \be \Gamma_\mathrm{a} ={2\pi \over \op }\sum_{n=0}^{\infty}\int^{\infty}_{-\infty}\int^t_0
{d (\Sigma d_n )\over dx}
 n^2b_n
\cos \chi_n(x,t')
dt'dx ,\ee
with $\chi_n(x,t')=3 n k_{0y}\op x (t - t')/2$, and 
 \be \Gamma_\mathrm{b}={2\pi \rp \over \op }\sum_{n=0}^{\infty}\int^{\infty}_{-\infty}\int^t_0
{d (\Sigma f_n )\over dx}
 n b_n
\sin \chi_n(x,t') 
dt'dx ,\ee
with $d_n = b_n -x\partial b_n/\partial x,$ and $f_n = (2/3\op)\partial^2 b_n/\partial x\partial t'.$

The expression for $\Gamma_\mathrm{b}$ can be integrated by parts with respect to $t'$ and combined with  that for $\Gamma_\mathrm{a}$ to yield the following expression for the total torque:
\be  \Gamma_\mathrm{c,lin} ={2\pi \over \op }\sum_{n=0}^{\infty}\int^{\infty}_{-\infty}\int^t_0{d \Sigma \over dx}n^2b_n^2\cos \chi_n(x,t')
dt' dx. \label{AppT}\ee

Equation (\ref{AppT}) shows how the corotation torque develops with time
after a protoplanet is inserted at $t=0.$ We shall see that this is on the time scale
of the orbital period. Consider first the case when the $b_n$ are independent
of time as would be the case for a large softening length
compared to the scale height  where there is  negligible pressure
response. Then the integral with respect to  $t'$ can be performed with the result that

\be  \Gamma_\mathrm{c,lin} ={4 \pi \rp\over 3\op^2 }\int^{\infty}_{-\infty}\sum_{n=0}^{\infty}
 ng_n(x=\zeta L_y/(3n\pi\op t))
{\sin\zeta\over \zeta} d\zeta,\ee
where $g_n(x) = b_n(x)^2(d\Sigma /dx).$

We remark that
most of the contribution to the integral comes
from $\zeta \sim 1.$ The corresponding value of $x  \sim L_y/(3n\pi\op t)
\sim 2 \rp /(3n\op t).$ Thus at early times the
 corotation torque has contributions
from mainly large radii but as time progresses the contributing region contracts
towards the corotation circle. To estimate the time involved, we note that
for large softening,  the  value
of $n$ giving the dominant Fourier components for the corotation torque 
is expected to be $n \sim 1/b,$
with  contributions from larger values of $n$ being reduced because of smoothing.
We then see that the region contributing to the corotation torque
is within $b\rp$ of the corotation circle within an orbital period 
independently of its size.
Thus the linear corotation torque in this model is established in about an orbital period.

The limiting value is easily obtained by letting 
$t \rightarrow \infty$ in the above expression.
The result after performing the integration is

\be  \Gamma_\mathrm{c,lin} ={4 \pi^2 \rp\over 3\op^2 }\sum_{n=0}^{\infty}
 nb_n(0)^2(d\Sigma /dx)_{x=0}\label{APPR}
.\ee
This is exactly what is obtained from the expression
(\ref{eqTmlin}) given by
\citet{1979ApJ...233..857G}  recalling  that the vorticity in the
background local model is constant and equal to $\op/2$
making the vortensity gradient proportional to the gradient of $1/ \Sigma.$ 
To illustrate the density response in the neighbourhood
of corotation we plot the linear surface density response
  for the case with $\alpha=-2$, $b=0.03$, $\hp=0.05$
  and $q=1.26\cdot 10^{-5}$
 in figure
\ref{figlindens}. This was 
 calculated following the procedure described in section
\ref{secLin}. In addition to the strong wake produced by the Lindblad
torques there is both  an overdensity leading and an underdensity trailing
 the planet. These narrow features are localized on the corotation
circle as expected from the analysis presented here. 
These produce a positive torque acting on the planet.

Although (\ref{APPR}) has been obtained 
 assuming the softening length ls large compared to the scale height, we expect it to apply more generally.
When the softening parameter is not large and the generalised potential, $\Phi_\mathrm{Gp}$  is used, the added enthalpy perturbation results in the $b_n$ being unknown functions of time.
However, the linear response calculation of \citet{1979ApJ...233..857G}
indicates that the scale of the response is  $|x| \sim H,$
with the dominant values of $n \sim \rp/H.$
Equation (\ref{AppT}) then indicates only values of $|t-t'| < \sim \rp/(nH\op)  \sim
\op^{-1}$ will contribute significantly to the torque.
Thus when $\Phi_\mathrm{Gp}$ varies slowly compared to the orbital period, 
expected as the linear response approaches its steady value, $b_n$ may be taken
to be locally constant in time, and
then from the above discussion,  the corotation torque will be the local \citet{1979ApJ...233..857G} value.  The situation is also similar at early times.
For $ t < \sim \rp/(nH\op) \sim
\op^{-1},$ the cosine term in equation (\ref{AppT}) may be replaced by unity.
This then implies that at early times  
 \be \Gamma_\mathrm{c,lin} ={4\pi^2 \rp\over 3\op^2 }\sum_{n=0}^{\infty}
{3nH\op t \over \pi  \rp}\left \langle{d \Sigma \over dx}
 nb_n^2\right\rangle,\label{AppT1}\ee
where the angled brackets denote an appropriate integral mean.
When $t$ approaches $\op^{-1},$ this becomes the same mean of 
torques of the form (\ref{eqTmlin}).

Note that  
the time to establish the linear response and corotation torque, the latter 
expected to be finite, does not depend
on the size of the perturbation, or accordingly,
$q.$ This time should be an intrinsic time
which can only be a multiple of the orbital period.  Accordingly for
  sufficiently small $q$ we can  ensure that the system is in the
linear regime when the torque is set up so that the linear formulation should be valid
then.

\end{document}